\documentclass[showpacs,amsmath,amssymb,twocolumn,floatfix,prl]{revtex4}
\usepackage[dvips]{graphicx}

\def\calh{{\mathcal H}}
\def\calp{{\mathcal P}}
\def\ket#1{|#1\rangle}
\def\bra#1{\langle#1|}
\def\braket#1#2{\langle #1 | #2 \rangle}
\def\tr{\textrm{tr}}

\def\g{\gamma}

\begin{document}

\title{Entropy of entanglement and multifractal exponents for random states}
\author{Olivier Giraud, John Martin and Bertrand Georgeot}
\affiliation{Laboratoire de Physique Th\'eorique, Universit\'e de
Toulouse, UPS, CNRS, 31062 Toulouse France}
\date{\today}

\begin{abstract}
We relate the entropy of entanglement of ensembles of random
vectors to their generalized fractal dimensions. Expanding the von
Neumann entropy around its maximum we show that the first order only
depends on the participation ratio, while higher orders involve
other multifractal exponents. These results can be applied to
entanglement behavior near the Anderson transition.
\end{abstract}
 \maketitle

Entanglement is an important characteristics of quantum systems,
which has been much studied in the past few years due to its
relevance to quantum information and computation. It is a feature
that is absent from classical information processing, and a crucial
ingredient in many quantum protocols. In the field of quantum
computing, it has been shown that a process involving pure states
with small enough entanglement can always be simulated efficiently
classically~\cite{jozsa}. Thus a quantum algorithm exponentially
faster than classical ones requires a minimal amount of entanglement
(at least for pure states). Conversely, it is possible to take
advantage of the weak entanglement in certain quantum many-body
systems to devise efficient classical algorithms to simulate them
~\cite{cirac}. All these reasons make it important to estimate the amount of
entanglement present in different types of physical systems, and
relate it to other properties of the system. However, in many cases
the features specific to a system obscure its generic behavior. One
way to circumvent this problem and to extract generic properties is
to construct ensembles of systems which after averaging over random
realizations can give analytic formulas. Such an approach has proven
successful, e.g.\ in the quantum chaos field, where Random Matrix
Theory (RMT) can describe many properties of complex
quantum systems.

One of the interesting questions which have been addressed in many
studies (see e.g.~\cite{spins} and references therein) is the
behavior of entanglement near phase transitions. It has been
shown that the entanglement of the ground state changes close to
phase transitions. For example, in the {\it XXZ} and {\it XY} spin chain
models, the entanglement between a block of spins and the rest of
the system diverges logarithmically with the block size at the
transition point~\cite{kitaev}, making classical simulations harder. 
However, such results cannot be applied
directly to systems where the transition concerns one-particle
states, for which entanglement has to be suitably defined. A famous
example is the Anderson transition of electrons in a disordered
potential, which separates localized from extended states, with
multifractal states at the transition point. Previous works
\cite{oneparticle} have described the lattice on which the particle evolves
as a spin chain and studied entanglement in this framework. However, 
the lattice can alternatively be described in terms of quantum 
computation with a much smaller number of two-level systems~\cite{pom}. 

In this paper, we study entanglement of random vectors which can be
localized, extended or multifractal in Hilbert space. We consider
entanglement between blocks of qubits. In the case of the Anderson
transition, this amounts to directly relate entanglement to the
quantum simulation of the system on a $n_r$--qubit system, the
number of lattice sites being $2^{n_r}$ rather than $n_r$ as in
\cite{oneparticle}.
Entanglement of random pure states was mainly studied in the case of
columns of matrices drawn from the Circular Unitary Ensemble (CUE)
\cite{CUE}. However, such vectors are extended and cannot describe
systems with various amounts of localization, from genuine
localization to multifractality. Recently it was shown in
\cite{GirMarGeo,viola} that for localized random vectors, the linear
entanglement entropy (first order of the von Neumann entropy)
of one qubit with all the others can be related to the
localization properties.
Here we develop this approach to obtain a general
description of bipartite entanglement in terms of certain global
properties for random vectors both extended and localized. First we
show that for any bipartition, the linear
entropy can be written in
terms of the participation ratio, a measure of localization. We then
show that higher-order terms also depend on higher moments of the
wavefunction. In particular, for multifractal systems they are
controlled by the multifractal exponents.

Bipartite entanglement of a pure state $|\psi\rangle$ belonging to a
Hilbert space $\calh_A\otimes\calh_B$ is measured through the
entropy of entanglement, which has been shown to be a unique
entanglement measure \cite{PopRoh}. Let $\rho_A$ be the density
matrix obtained by tracing subsystem $B$ out of
$\rho=\ket{\psi}\bra{\psi}$. The entropy of entanglement of the
state with respect to the bipartition $(A,B)$ is the von Neumann
entropy of $\rho_A$, that is $S=-\tr(\rho_A\log_2\rho_A)$. It is
convenient to define the linear entropy as
$S_L=\frac{d}{d-1}(1-\tr\rho_A^2)$, where $d=\dim\calh_A\leq
\dim\calh_B$. The scaling factor ensures that $S_L$ varies in
$[0,1]$. We will show that the average value of $S$ over a set of
random states can be expressed only in terms of averages of the
moments of the wavefunction
\begin{equation}\label{moments}
p_q=\sum_{i=1}^{N}|\psi_i|^{2q}
\end{equation}
provided some natural assumptions are made. Here we consider ensembles of
random vectors of size $N\equiv 2^{n_r}$ with the following two
properties: i) the phases of the vector components are independent,
uniformly distributed random variables, and ii) the joint
distribution $P(x_1,\ldots,x_N)$ of the modulus squared 
of the vector components is such that all marginal distributions $P(x_i)$,
$P(x_i,x_j)$ for $i\neq j$, $P(x_i,x_j,x_k)$ for $i\neq j\neq k$ and
so on, do not depend on the indices. As a consequence, all
correlators
$\langle|\psi_{i_1}|^{2s_1}|\psi_{i_2}|^{2s_2}\ldots\rangle$ of the
components of $\ket{\psi}$ are independent of the indices $i_1\ne i_2\ne
\ldots$ involved. Random vectors realized as columns of CUE matrices
are instances of vectors having such properties.

Let us first consider the simplest case of entanglement of one qubit
with respect to the others. Then $d=2$ and the linear entropy $S_L$
is simply the {\it tangle} $\tau$, or the square of the generalized
concurrence~\cite{RunCav}. It is given by $\tau=4\det\rho_A$. If we
consider a vector $\ket{\psi}$ of size $N$, the bipartition with
respect to qubit $i$ splits the components $\psi_j$ of $\ket{\psi}$
into two sets, according to the value of the $i$th bit of the binary
decomposition of $j$. If $\ket{\psi^{(0)}}$ and $\ket{\psi^{(1)}}$
are the two corresponding vectors, the linear entropy is
\begin{equation}
\label{tau}
\tau=4\left(\braket{\psi^{(0)}}{\psi^{(0)}}\braket{\psi^{(1)}}{\psi^{(1)}}-|\braket{\psi^{(0)}}{\psi^{(1)}}|^2\right).
\end{equation}
After averaging $\tau$ over random phases, only the diagonal terms
survive in the scalar product $|\braket{\psi^{(0)}}{\psi^{(1)}}|^2$.
Since it is assumed that two-point correlators of the vector
$\ket{\psi}$ do not depend on indices, their average can be
expressed solely in terms of the mean moments, as
$\langle|\psi_i|^2|\psi_j|^2\rangle=\frac{\langle
p_1^2\rangle-\langle p_2\rangle}{N(N-1)}$ for $i\ne j$.
Normalization of $\ket{\psi}$ implies $p_1=1$. As vectors
$\ket{\psi^{(0)}}$ and $\ket{\psi^{(1)}}$ always contain components
of $\ket{\psi}$ with different indices, we get
\begin{equation}
\label{tau1} \langle\tau\rangle=\frac{N-2}{N-1}(1-\langle
p_2\rangle).
\end{equation}
Since $p_2=1/\xi$ where $\xi$ is the inverse participation ratio
(IPR), Eq.~\eqref{tau1} is exactly the Eq.~(3) of
Ref.~\cite{GirMarGeo}. Let us now turn to the general case, and
consider the entropy of entanglement of $\nu$ qubits with $n_r-\nu$
others ($(\nu, n_r-\nu)$ bipartition). The vector $\ket{\psi}$ is
now split into vectors $\ket{\psi^{(j)}}$, $0\leq j\leq 2^{\nu}-1$,
depending on the values of the $\nu$ qubits. The reduced density
matrix $\rho_A$ then appears as the Gram matrix of the
$\ket{\psi^{(j)}}$, and the linear entropy is
\begin{equation}
\label{taugeneralth} S_L=\frac{2^{\nu}}{2^{\nu}-1}\left(1
-\sum_{i,j=0}^{2^{\nu}-1}|\braket{\psi^{(i)}}{\psi^{(j)}}|^2\right).
\end{equation}
When averaging over random vectors, each term in
Eq.~(\ref{taugeneralth}) with $i\neq j$ yields $2^{n_r-\nu}$
two-point correlators, while each term with $i=j$ yields
$2^{n_r-\nu}(2^{n_r-\nu}-1)$ two-point correlators and $2^{n_r-\nu}$
terms of the form $\langle|\psi_i|^4\rangle$. Inserting these
expressions into \eqref{taugeneralth} gives
$\langle S_L\rangle=(N-2^{\nu})(1-\langle p_2\rangle)/(N-1)$, 
which generalizes Eq.~\eqref{tau1}. The
first-order series expansion of the mean von Neumann entropy around its
maximum can be expressed as
\begin{equation}\label{SSL}
\langle S\rangle\simeq \nu-\frac{2^\nu-1}{2\ln 2}\left(1-\frac{N-2^{\nu}}{N-1}
\left(1-\langle p_2\rangle\right)\right),
\end{equation}
with $p_2=1/\xi$. Equation~\eqref{SSL} shows that for any partition
of the system into two subsystems, the average bipartite
entanglement of random states only depends at first order on the
localization properties of the states, through the mean
participation ratio. For CUE vectors, formula~(\ref{SSL})
reduces to the expression for the mean entanglement derived earlier
in~\cite{Sco}. More interestingly, this formula also applies to
multifractal quantum states. There, the asymptotic behavior of the
IPR is governed by the fractal exponent $D_2$, where one defines
generalized fractal dimensions $D_q$ through the scaling of the
moments $p_q\propto N^{-D_q(q-1)}$. Thus the linear entropy is only
sensitive to a single fractal dimension. These results imply that
entanglement grows more slowly with the system size for multifractal
systems.

\begin{figure}
\begin{center}
\includegraphics[width=.95\linewidth]{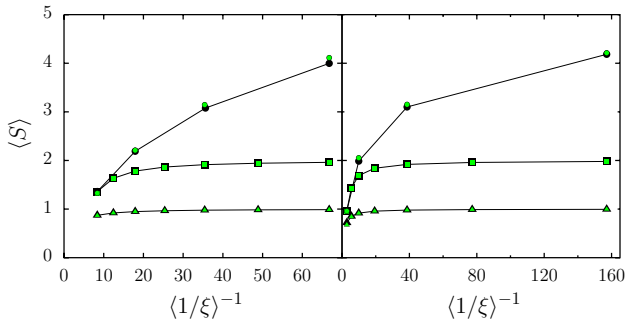}
\end{center}
\caption{(Color online) Mean entropy of entanglement as a
function of the mean IPR. Left: eigenvectors of (\ref{ISRM}) with
$\g=1/3$; the average is taken over $10^6$ eigenvectors. Right:
eigenvectors of the Hamiltonian $H$ (see text) with $\delta
=\Delta_0$ and $J/\delta=1.5$; average over $N/16$ central
eigenstates, with a total number of vectors $\approx 3\times 10^5$.
Triangles correspond to $\nu=1$, squares to $\nu=2$ and circles to
$\nu=n_r/2$, with $n_r=4-10$ (bipartition of the $\nu$ first qubits
with the $n_r-\nu$ others). Black symbols are the theoretical
predictions for $\langle S\rangle$ at first order (Eq.~\eqref{SSL}) 
and green (grey) symbols are the computed mean values of the exact 
$\langle S\rangle$. }
\label{firstorder}
\end{figure}

To test the relevance of Eq.~\eqref{SSL} for describing entanglement
in realistic settings, 
we consider eigenvectors of $N\times N$ unitary matrices of the form
\begin{equation}\label{ISRM}
U_{kl}=\frac{e^{i\phi_k} }{N}\frac{1-e^{2i\pi N \g}}{1-e^{2i\pi (k-l+N\g)/N}},
\end{equation}
where $\phi_k$ are independent random variables uniformly distributed in $[0,
2\pi[$. These random matrices display intermediate statistical
properties~\cite{bogomolny}, and possess eigenvectors that are
multifractal~\cite{MarGirGeo}, both features being tuned through the
value of the real parameter $\g$. We also illustrate
Eq.~\eqref{SSL} with eigenstates of a many-body
Hamiltonian with disorder and interaction $H = \sum_{i} \Gamma_i
\sigma_{i}^z + \sum_{i<j} J_{ij} \sigma_{i}^x \sigma_{j}^x$. This
system can describe a quantum computer in presence of static
disorder \cite{qchaos}. Here the $\sigma_{i}$ are the Pauli matrices
for qubit $i$, energy spacing between the two states of qubit $i$ is
given by $\Gamma_i$ randomly and uniformly distributed in the
interval $[\Delta_0 -\delta /2, \Delta_0 + \delta /2 ]$, and
the $J_{ij}$ uniformly distributed in the interval $[-J,J]$ represent a
random static interaction. For large $J$  and $\delta \approx
\Delta_0$, eigenstates are delocalized in the basis of register states, 
but without multifractality. They display properties of quantum
chaos, with eigenvalues statistics close to the ones of RMT
\cite{qchaos}. For both systems (unitary matrices and many-body
Hamiltonian), components of the eigenvectors have been shuffled in
order to reduce correlations, but leaving the peculiarities of the
distribution itself unaltered. Figure~\ref{firstorder} plots the
first-order expansion (\ref{SSL}) as a function of the mean IPR for
three different bipartitions, showing remarkable agreement with the
exact $\langle S\rangle$, 
both for multifractal (Fig.~\ref{firstorder}, left panel)
and non fractal (right) states, and even for moderately entangled
states.  The agreement is better for the non-fractal system than for
the multifractal one.  This can be understood from the study of
higher order terms in the entropy.

Indeed, while the linear entropy does not depend on other fractal
dimensions than $D_2$, the entropy of entanglement does. If we go
back to the case of a $(1, n_r-1)$ bipartition of the system, the
entropy of entanglement can be expressed in a simple way as a
function of $\tau$ as
\begin{equation}\label{S}
S(\tau)=h\left(\frac{1+\sqrt{1-\tau}}{2}\right),
\end{equation}
where $h(x)=-x\log_2 x-(1-x)\log_2(1-x)$. The series expansion of
$S(\tau)$ up to order $m$ in $(1-\tau)$ reads
\begin{equation}
\label{expansionS} S_m(\tau)=1-\frac{1}{\ln
2}\sum_{n=1}^{m}\frac{(1-\tau)^n}{2n(2n-1)}.
\end{equation}
The tangle $\tau$ corresponds, up to a linear
transformation, to $S_1(\tau)$. Let us now calculate the average 
of higher orders in this expansion. The second-order expansion of $S(\tau)$
involves calculating the mean value of
\begin{equation}
\label{uv4}
|\braket{\psi^{(0)}}{\psi^{(1)}}|^4=\sum_{i,j,k,l=1}^{N/2} u_i^* u_j
u_k^* u_l v_i v_j^* v_k v_l^*,
\end{equation}
where the star denotes complex conjugation and $u_i$, $v_i$ are the
components of $\ket{\psi^{(0)}}$, $\ket{\psi^{(1)}}$ respectively.
Under the assumption of random phases, only terms whose phases
cancel survive in \eqref{uv4}. Since the phases of all components of
$\ket{\psi}$ are independent, cancelation of the phase can only
occur if the sets $\{i,k\}$ and $\{j,l\}$ are equal. Thus
\begin{equation}\label{uv4b}
\langle|\braket{\psi^{(0)}}{\psi^{(1)}}|^4\rangle=2\sum_{\genfrac{}{}{0pt}{}{i,
k=1}{i\neq k}}^{N/2} \langle |u_i u_k v_i v_k|^2\rangle
+\sum_{i=1}^{N/2}\langle|u_i v_i|^4\rangle.
\end{equation}
The correlators in Eq.~(\ref{uv4b}) can be expressed as a function
of the moments as follows. Using standard notations \cite{MacDo}, we
will denote by $\lambda\vdash n$ a partition
$\lambda=(\lambda_1,\lambda_2,\ldots)$ of $n$, with
$\lambda_1\geq\lambda_2\geq\ldots$. For any partition $\lambda\vdash
n$, we define $p_{\lambda}=p_{\lambda_1}p_{\lambda_2}\ldots$, where
$p_{\lambda_i}$ are given by (\ref{moments}). The monomial symmetric
polynomials are defined as $ m_{\lambda}=\sum
\langle|\psi_1|^{2\lambda_1}|\psi_2|^{2\lambda_2}\ldots\rangle$
(the sum runs over all $\mathcal{P}_{\lambda}$ permutations of the
$\lambda_i$), and we set
$c_{\lambda}=m_{\lambda}/\mathcal{P}_{\lambda}$. The $p_{\lambda}$
and $m_{\lambda}$ are related by the simple linear relation
$p_{\lambda}=\sum_{\mu}L_{\lambda\mu}m_{\mu}$, where
$L_{\lambda\mu}$ is an invertible integer lower-triangular matrix
(\cite{MacDo}, p.103). Upon our assumption ii), any correlator of
the form $\langle|\psi_{i_1}|^{2s_1}|\psi_{i_2}|^{2s_2}\ldots\rangle$ is
equal to a $c_{\lambda}$ for some partition $\lambda$ of $n$, and
thus can be expressed as a function of the moments. For instance the
two correlators in Eq.~(\ref{uv4b}) are respectively equal to
$c_{1111}$ and $c_{22}$. Treating similarly all terms involved in
$\tau^2$ gives
\begin{eqnarray}
\label{tau2}
\langle\tau^2\rangle&=&N(N-2)(N^2-6N+16)c_{1111}\\
&+&4N(N-2)(N-4)c_{211}+4N(N-2)c_{22}.\nonumber
\end{eqnarray}
This term involves the calculation of three correlators. Using the
relation between the $c_{\lambda}$ and $p_{\lambda}$ and the fact
that the vectors are normalized to one we get
\begin{eqnarray}\label{cs}
c_{22}&=&\frac{\langle p_2^2\rangle-\langle p_4\rangle}{N(N-1)},\
c_{211}=\frac{\langle p_2\rangle-\langle p_2^2\rangle-2\langle p_3\rangle+2\langle p_4\rangle}{N(N-1)(N-2)},\nonumber\\
c_{1111}&=&\frac{1-6\langle p_2\rangle+8\langle p_3\rangle+3\langle
p_2^2\rangle-6\langle p_4\rangle}{N(N-1)(N-2)(N-3)}.
\end{eqnarray}

The calculation of the general term $\langle\tau^n\rangle$ 
can be performed along the same
lines. Expanding \eqref{tau} we get
\begin{eqnarray}
\label{taun}
\tau^n&=&4^n\sum_{k=0}^{n}\binom{n}{k}(-1)^{n-k}\\
&\times&\left(\sum_{i=1}^{N/2}|u_i|^2\right)^k
\left(\sum_{i=1}^{N/2}|v_i|^2\right)^k
\left(\sum_{i,j}u_i^* v_i u_j v_j^*\right)^{n-k}.\nonumber
\end{eqnarray}
The expansion of $\left(\sum_{i,j}u_i^* v_i u_j v_j^*\right)^t$
contains products of the form $(u_{i_1}\ldots
u_{i_t})^*u_{j_1}\ldots u_{j_t}$. Only terms where the phases coming
from the $u_{i_k}^*$ compensate those coming from the $u_{j_k}$
survive when averaging over random phases. Thus we keep only terms
where $\{j_1, \ldots, j_t\}$ is a permutation of $\{i_1, \ldots,
i_t\}$. If $\calp_{K}$ is the number of permutations of a set $K$,
the average of \eqref{taun} over random vectors reads
\begin{eqnarray}
\label{eqhorr} \langle\tau^n\rangle=\Big\langle
4^n\sum_{k=0}^{n}\binom{n}{k}(-1)^{n-k}\sum_{\genfrac{}{}{0pt}{}{p_1,
\ldots, p_k}{q_1, \ldots, q_k}}
\prod_{j=1}^{k}|u_{p_j}|^2|v_{q_j}|^2\nonumber\\
\times \sum_{i_1,\ldots, i_{n-k}}\calp_{\{i_1,\ldots, i_{n-k}\}}
|u_{i_1}v_{i_1}|^2\ldots |u_{i_{n-k}}v_{i_{n-k}}|^2\Big\rangle.
\end{eqnarray}
Terms with the same correlator can be grouped together. Each
correlator in \eqref{eqhorr} is some $c_{\lambda\cup\lambda'}$, with
$\lambda,\lambda'$ partitions of $n$. For $\lambda\vdash n$ and
$\mu\vdash k$ we define the coefficient
$A_{\lambda\mu}=\frac{k!}{\mu!}\sum\frac{(n-k)!}{({\bf s-k})!}$,
where the sum runs over all vectors ${\bf s}=(s_1,\ldots,s_{N/2})$ 
which are permutations of $\lambda$, and
${\bf k}=(\mu_1,\ldots,\mu_{N/2})$. We have used the notations 
$\mathbf{a}=(a_1,a_2,\ldots)$ and ${\bf a}!=a_1!a_2!\ldots$. Finally we get
\begin{equation}
\label{taunfinal} \langle\tau^n\rangle
=4^n\sum_{\lambda,\lambda'\vdash n}\bigg(
\sum_{k=0}^{n}\binom{n}{k}(-1)^{k} \sum_{\mu\vdash k}\calp_{\mu}
A_{\lambda\mu}A_{\lambda'\mu} \bigg)c_{\lambda\cup\lambda'},
\end{equation}
which provides an expression for the $n$th
order for $\langle S\rangle$ 
as a function of the $\langle p_\lambda\rangle$. In the
special case of CUE random vectors, by resumming the whole series we
recover after some algebra the well-known result ~\cite{Pag}
$\langle S(\tau)\rangle=\frac{1}{\ln
2}\sum_{k=N/2+1}^{N-1}\frac{1}{k}$.
Note that similar expressions can be derived for a general $(\nu,
n_r-\nu)$ bipartition. In this case, the entropy
$S=-\tr(\rho_A\log_2\rho_A)$ can be expanded around the maximally
mixed state $\rho_0=\mathbf{1}/2^{\nu}$, as
\begin{equation}
\label{Snu} S=\nu+\frac{1}{\ln 2}\sum_{n=1}^{\infty}
\frac{(-2^\nu)^{n}}{n(n+1)}\tr((\rho_A-\rho_0)^{n+1}).
\end{equation}
After averaging over random vectors, one can check that the traces
in \eqref{Snu} can be written as
$\langle\tr\rho_A^k\rangle=\sum_{\lambda\vdash
  k}a^{(k)}_{\lambda}c_{\lambda}$,
with $a^{(k)}_{\lambda}$ some integer combinatorial coefficient. The
entropy can thus be written as a linear combination of $c_{\lambda}$
with rational coefficients that can be expressed in terms of the
$a^{(k)}_{\lambda}$.

\begin{figure}
\begin{center}
\includegraphics[width=.95\linewidth]{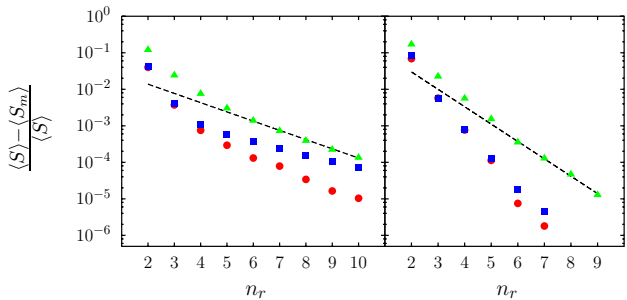}
\end{center}
\caption{(Color online) Relative difference of the entropy of
entanglement (\ref{S}) and its successive approximations $S_m$
($m=1,2$) with respect to the number of qubits for eigenvectors of
(\ref{ISRM}) for (left) $\g=1/3$ and (right) $\g=1/7$. The average
is taken over $10^7$ eigenvectors, yielding an accuracy $\lesssim
10^{-6}$ on the computed mean values. Green triangles correspond to
the first-order expansion $S_1$, blue squares and red circles to the
second-order expansion $S_2$. The difference between the latter two
is that for blue squares $\langle p_2^2 \rangle$ appearing in
Eq.~(\ref{cs}) has been replaced by $\langle p_2 \rangle^2$ yielding
a less accurate approximation. Dashed line is a linear fit yielding
$1-\langle S_1\rangle/\langle S\rangle$ $\sim N^{-0.84}$ for
$\g=1/3$ and $N^{-1.58}$ for $\g=1/7$. } \label{secondorder}
\end{figure}

In Fig.~\ref{secondorder} we illustrate the accuracy of higher-order
terms in the series expansion of $S$ for multifractal random vectors by
comparing the first and second-order expansion 
for eigenvectors of the matrices (\ref{ISRM}). As
expected, the second-order expansion is much more accurate than the
first order one and gives a much better estimate of the mean entropy
of entanglement already for small system sizes. For large $N$, the
dominant term in $S_2$ is $\propto\langle p_2^2\rangle$. Numerically
we obtained $\langle p_2^2\rangle\sim N^{-0.81}$ for $\g=1/3$, and
$\langle p_2^2\rangle\sim N^{-1.53}$ for $\g=1/7$, which is indeed
consistent with the slopes of the linear fit of $\log_2 (1-\langle
S_1\rangle/\langle S\rangle)$ (see Fig.~\ref{secondorder}). If one
replaces $\langle p_2^2 \rangle$ appearing in Eq.~(\ref{cs}) by
$\langle p_2 \rangle^2$ (squares in Fig.~\ref{secondorder}), the
second-order expansion is now governed only by three multifractal
dimensions $D_2$, $D_3$, $D_4$. Although it becomes less and less
accurate with the system size because of the increase of the
variance of $p_2$, it remains a very good improvement over the first
order in the case of moderate multifractality
(Fig.~\ref{secondorder}, right).

Our results show that the entanglement of random vectors directly 
depends on whether they are localized, multifractal
or extended. The numerical simulations for different physical examples
show that our theory describes well individual systems whose
correlations are averaged out. Previous results \cite{GirMarGeo}
have shown that Anderson-localized states have entanglement going to
zero for large system size. The present work shows that 
multifractal states, such as those appearing at the Anderson transition,
approach the maximal value of entanglement in a way controlled by the
multifractal exponents. Although
extended and multifractal states are both close to maximal
entanglement, the way multifractal states approach the maximal value
for large system size is slower.

The authors thank CalMiP in Toulouse for access to their
supercomputers. This work was supported by the Agence Nationale de
la Recherche (ANR project INFOSYSQQ, contract number
ANR-05-JCJC-0072) and the European program EC IST FP6-015708
EuroSQIP.

\end{document}